# Online reduced-order modeling from primal solver trajectories for PDE-constrained optimization

Zhe Lu [a,b,c], Wenbo Cao [d,e], Weiwei Zhang [a,b,c,*]

[a] *School of Aeronautics, Northwestern Polytechnical University, Xi'an 710072, China*

[b] *International Joint Institute of Artificial Intelligence on Fluid Mechanics, Northwestern Polytechnical University, Xi'an 710072, China*

[c] *National Key Laboratory of Aircraft Configuration Design, Xi'an 710072, China*

[d] *Institute of AI for Industries, Chinese Academy of Sciences, Nanjing 211135, China*

[e] *Institute of Computing Technology, Chinese Academy of Sciences, Beijing 100190, China*

* *Corresponding author. E-mail address: aeroelastic@nwpu.edu.cn*

**Abstract.** Repeated high-fidelity state and gradient evaluations make large-scale PDE-constrained optimization computationally expensive. Reduced-order modeling provides a natural way to reduce the cost of these repeated high-dimensional computations. However, existing reduced-order modeling approaches often rely on converged high-fidelity states for model construction or updating, while their reduced spaces may lose accuracy as the design moves beyond the region represented by the available data. To address these limitations, we present Reduced-order Trajectory-based Optimization (RETO), an online framework that constructs local reduced-order models (ROMs) directly from unconverged primal solver trajectories. Each local ROM is built from snapshots collected during a finite number of high-fidelity primal iterations, without requiring full-order convergence, and supports multiple low-cost state evaluations, adjoint-based gradient evaluations, and design updates. New solver trajectories are generated at updated design points to reconstruct the local ROM and continue the optimization. Across three distinct PDE-constrained optimization problems, RETO achieves solution quality comparable to that of full-order optimization, with speedups of approximately 8–22× in cases without discontinuities in the state fields. Even in the transonic inverse-design case involving moving shocks, RETO retains an approximately threefold speedup. The transonic results further show that the online local ROM progressively captures evolving shock-related flow features as new primal solver trajectories are incorporated. These results demonstrate the value of unconverged primal solver trajectories as an online information source for reduced-order optimization, allowing the reduced basis to evolve with the design rather than requiring a basis that covers the entire design space in advance.

## 1. Introduction

PDE-constrained optimization is widely used in parameter inversion, optimal control, aerodynamic design, and other simulation-based engineering optimization problems [1, 2]. Its central computational task is to repeatedly solve the high-fidelity state equations as the design variables are updated and to evaluate the corresponding objective gradients. For large-scale discrete systems, particularly those with high-dimensional state variables and strong nonlinearities, even a single high-fidelity state solution and gradient evaluation can be expensive. These calculations must then be repeated at successively updated design points throughout the optimization. Adjoint methods substantially reduce the cost of gradient evaluation when the number of design variables is large [3]. Their main advantage is that the sensitivity cost does not grow linearly with the number of design variables. They do not, however, eliminate the need to solve the high-dimensional state equations and the associated linearized systems at each design point. Reducing the computational cost of these repeated state and gradient evaluations while retaining the high-fidelity physical constraints and reliable gradient information therefore remains an important computational issue in large-scale PDE-constrained optimization.

Reduced-order modeling provides a well-established route for reducing the repeated high-dimensional computations encountered in simulation, control, and optimization [4-6]. Over the past several decades, reduced-order models, particularly those based on proper orthogonal decomposition (POD), have been applied to PDE-constrained optimal control, fluid-flow control, and aerodynamic design to replace part of the repeated high-fidelity computation with low-dimensional approximations [7-9]. In such approaches, the state response is represented in a low-dimensional trial space whose dimension is much smaller than that of the full-order system, and the corresponding state and gradient calculations can therefore be performed at substantially reduced cost. However, the effectiveness of a reduced-order model depends strongly on whether its reduced space adequately represents the states encountered during optimization. As the optimization trajectory departs from the region or reference trajectory represented by the training data, a fixed reduced space may progressively lose accuracy or validity [10, 11]. Expanding the offline training region, adding high-fidelity samples, or increasing the reduced dimension can improve model coverage, but these measures also increase offline cost or online complexity. Moreover, an optimization process typically explores only a limited portion of the parameter space. Constructing a globally accurate reduced model over the entire admissible design domain may therefore require substantially more high-fidelity information than is actually needed along the optimization trajectory. This motivates the use of local information acquired along the optimization path to update the reduced model for the current design neighborhood.

A number of local reduced-basis and online updating strategies have been developed to address the limited validity of a fixed reduced space during optimization. Amsallem et al. constructed local reduced bases so that the low-dimensional representation could adapt to different regions of parameter space [12]. Zahr and Farhat progressively constructed the reduced model along a PDE-constrained optimization trajectory, allowing high-fidelity information to be incorporated as the optimization proceeded [10]. Li et al. proposed a continually moving ROM in which the local model moves and is updated during aerodynamic optimization [13]. Wen and Zahr combined on-the-fly model construction with hyperreduction and trust-region model management for large-scale shape optimization [14]. Qian et al. coupled reduced basis models with trust-region management and used a posteriori estimates of the objective and gradient errors to control the use and update of the reduced model [15]. These studies show that continually adding local high-fidelity information along the optimization trajectory can maintain the validity of the reduced model in the current design region more directly than constructing a single global model over the entire parameter domain in advance. Such approaches may still require newly converged high-fidelity states when the reduced model is updated. Consequently, the cost of repeatedly advancing the high-fidelity primal solver to convergence is not eliminated and can remain substantial in online reduced-order optimization.

The intermediate states naturally generated during high-fidelity primal advancement provide another source of information that can be exploited. As the primal solver advances from its initial state toward a steady solution, it produces a sequence of unconverged states. These states are normally not retained as final physical solutions, but they contain information about the evolution of the iterative process. Previous studies of iterative solver acceleration have shown that such intermediate information need not be treated merely as a computational by-product. Methods such as Anderson acceleration exploit information from previous iterates and residuals to accelerate nonlinear iterative solvers [16]. In CFD, unconverged primal states have also been used to construct low-dimensional models that generate lower-residual states, thereby accelerating convergence and stabilizing iterative processes that would otherwise have difficulty reaching a steady solution [17, 18]. More recent work on general iterative solvers has likewise shown that intermediate information generated during the solution process can be reused to reduce the computational complexity of repeated solves [19]. Although these studies focus on the solution process itself rather than online reduced-order modeling for PDE-constrained optimization, they support an important observation: useful information from a high-fidelity computation is not confined to the final converged solution, and intermediate iterative states can also be reused. These observations motivate the direct use of primal iterative trajectories in PDE-constrained optimization, so that information from one finite high-fidelity advancement can support not only the current design point but also several subsequent state, gradient, and design updates.

Based on this idea, we develop a Reduced-order Trajectory-based Optimization (RETO) method. RETO does not require a newly converged high-fidelity state at every design update. Instead, the primal solver is advanced for a finite number of iterations at the current design point, and the unconverged states naturally generated during this process are used to construct a local reduced model for state and gradient evaluation. This model then supports several successive low-cost design updates.

As the optimization trajectory evolves, another finite high-fidelity advancement supplies new local state information and the reduced model is rebuilt. High-fidelity information acquisition and reduced-order optimization thus alternate along the design trajectory.

The main contributions are threefold. First, we develop a PDE-constrained optimization framework that uses unconverged iterative trajectories directly for online local reduced-order modeling, so that construction of a new local model does not require a newly converged high-fidelity solution. Second, the trajectory-based local state representation is combined with reduced-order adjoint gradient evaluation, allowing a single high-fidelity advancement to support multiple subsequent state evaluations, gradient evaluations, and design updates before the local model is rebuilt. Third, RETO is assessed using a Poisson parameter field inversion problem, a logistic resource allocation problem, and subsonic and transonic RANS airfoil pressure inverse-design problems. These cases examine solution quality, computational efficiency, and the evolution of the local model along the optimization trajectory. Table 1 compares RETO with representative online and local reduced-order optimization methods in terms of the information used for model construction, the state solution, and the optimization mechanism.

Table 1. Comparison of representative online/local reduced-order optimization methods with RETO

| | **Online model construction** | | **Optimization mechanism** | **Key characteristic** |
|---|---|---|---|---|
| | **Information source** | **State solution** | | |
| Zahr-Farhat [10] | High-fidelity states + sensitivities along optimization trajectory | Nonlinear LSPG | ROM gradient + trust region | Progressive ROM with rigorous model management |
| Li et al. [13] | Neighboring designs + moving reference states | Local first-order response model | Genetic algorithm | Continually moving ROM |
| Yao et al. [20] | State + sensitivity + residual-based enrichment trigger | LSPG + Newton | ROM gradient + trust-region optimization | Enriched state/sensitivity information |
| Wen-Zahr [14] | High-fidelity states + residual information | Hyperreduced ROM | Trust-region optimization | On-the-fly hyperreduction |
| **RETO** | **Unconverged primal trajectories + recent history** | **Local full-residual least-squares correction** | **Reduced-order adjoint gradient optimization** | **One high-fidelity acquisition supports multiple state, gradient, and design updates** |

Note: ROM, reduced-order model; LSPG, least-squares Petrov–Galerkin.

## 2. Reduced-order Trajectory-based Optimization

### 2.1 PDE-constrained optimization and adjoint gradient evaluation

Consider a steady PDE-constrained optimization problem after spatial discretization, written in the general form

$$\min_{\boldsymbol{\mu}\in\mathcal{D}} J(\mathbf{y},\boldsymbol{\mu}) \quad \text{s.t.} \quad \mathbf{R}(\mathbf{y},\boldsymbol{\mu})=\mathbf{0} \tag{1}$$

Here, $\mathbf{y}\in\mathbb{R}^{N_y}$ denotes the discrete state variables, $\boldsymbol{\mu}\in\mathbb{R}^{n_\mu}$ the design variables, $\mathbf{R}:\mathbb{R}^{N_y}\times\mathbb{R}^{n_\mu}\to\mathbb{R}^{N_R}$ the high-fidelity discrete PDE residual, $J:\mathbb{R}^{N_y}\times\mathbb{R}^{n_\mu}\to\mathbb{R}$ the optimization objective, and $\mathcal{D}$ the feasible set of the design variables. For the standard discrete state equations considered here, $N_R=N_y$. Bounds, equality constraints, and inequality constraints on the design variables can be handled separately by the corresponding optimizer. To focus on the structure of the state and gradient calculations, these algebraic constraints are not written explicitly in the following derivation.

For a given design $\boldsymbol{\mu}$, the full-order state $\mathbf{y}(\boldsymbol{\mu})$ is determined by the high-fidelity primal problem

$$\mathbf{R}(\mathbf{y},\boldsymbol{\mu})=\mathbf{0} \tag{2}$$

Solving Eq. (1) with a gradient-based method requires the derivative of the objective with respect to the design variables. In this work, the discrete adjoint method is used for gradient evaluation [3]. Its main advantage is that the objective gradient can be obtained without solving a separate state-sensitivity problem for each design variable, so that the gradient-evaluation cost does not grow linearly with the number of design variables.

Treating the state as an implicit function of the design variables, the total derivative of the objective can be written as

$$\frac{\mathrm{d}J}{\mathrm{d}\mu}=\frac{\partial J}{\partial\boldsymbol{\mu}}+\frac{\partial J}{\partial\mathbf{y}}\frac{d\mathbf{y}}{d\boldsymbol{\mu}} \tag{3}$$

Differentiating the state equation with respect to the design variables gives

$$\frac{\partial\mathbf{R}}{\partial\mathbf{y}}\frac{d\mathbf{y}}{d\boldsymbol{\mu}}+\frac{\partial\mathbf{R}}{\partial\boldsymbol{\mu}}=\mathbf{0} \tag{4}$$

Direct evaluation of the state sensitivities from Eq. (4) requires the corresponding linearized state equations to be solved for the design variables, and the computational cost generally grows with $n_{\boldsymbol{\mu}}$. To avoid explicitly forming $d\mathbf{y}/d\boldsymbol{\mu}$, we introduce an adjoint variable $\boldsymbol{\lambda}\in\mathbb{R}^{N_R}$ satisfying the discrete adjoint equation

$$\left(\frac{\partial \mathbf{R}}{\partial \mathbf{y}}\right)^T \boldsymbol{\lambda} = -\left(\frac{\partial J}{\partial \mathbf{y}}\right)^T \tag{5}$$

The gradient of the objective with respect to the design variables is then

$$\frac{\mathrm{d}J}{\mathrm{d}\boldsymbol{\mu}} = \frac{\partial J}{\partial \boldsymbol{\mu}} + \boldsymbol{\lambda}^T \frac{\partial \mathbf{R}}{\partial \boldsymbol{\mu}} \tag{6}$$

A single adjoint solution provides the objective gradient with respect to all design variables. However, repeated solution of the high-dimensional state and adjoint systems remains computationally expensive as the design changes. Reduced-order modeling is therefore introduced to lower the cost of these repeated evaluations.

### 2.2 Reduced-order state and gradient evaluation

To reduce the computational cost of repeated state and gradient evaluations, the high-dimensional state is approximated within a low-dimensional affine space,

$$\mathbf{y} \approx \bar{\mathbf{y}} + \boldsymbol{\Phi}\mathbf{a}, \quad \boldsymbol{\Phi} \in \mathbb{R}^{N_y \times r}, \quad \mathbf{a} \in \mathbb{R}^r, \quad r \ll N_y \tag{7}$$

Here, $\bar{\mathbf{y}}$ is the reference state, $\boldsymbol{\Phi}$ is an orthonormal reduced basis, and $\mathbf{a}$ contains the reduced coordinates. The reference state and basis are treated as fixed during each local reduced-order optimization stage. Their construction and updating are described in Section 2.3.

#### 2.2.1 Reduced-order state solution

Substituting Eq. (7) into the discrete residual gives $\mathbf{R}_\Phi(\mathbf{a}, \boldsymbol{\mu}) = \mathbf{R}(\bar{\mathbf{y}} + \boldsymbol{\Phi}\mathbf{a}, \boldsymbol{\mu})$. In the POD–Galerkin formulation, the reduced state is determined by requiring the residual to be orthogonal to the trial space,

$$\mathbf{R}_G(\mathbf{a}, \boldsymbol{\mu}) = \boldsymbol{\Phi}^T \mathbf{R}_\Phi(\mathbf{a}, \boldsymbol{\mu}) = \mathbf{0} \tag{8}$$

This formulation reduces both the state variables and residual equations to dimension $r$. However, the Galerkin condition does not directly constrain residual components outside the selected test space. To retain the full discrete residual in the state calculation, we instead primarily employ least-squares Petrov–Galerkin (LSPG) minimization,

$$\mathbf{a}^* = \arg\min_{\mathbf{a} \in \mathbb{R}^r} \frac{1}{2} \left\| \mathbf{R}_\Phi(\mathbf{a}, \boldsymbol{\mu}) \right\|_2^2 \tag{9}$$

Unlike the Galerkin formulation, LSPG reduces the dimension of the state search while retaining the original residual norm as the minimization criterion [21]. Defining the Jacobian with respect to the reduced coordinates as

$$\mathbf{A} = \frac{\partial \mathbf{R}_{\Phi}}{\partial \mathbf{a}} = \frac{\partial \mathbf{R}}{\partial \mathbf{y}} \boldsymbol{\Phi} \in \mathbb{R}^{N_R \times r} \tag{10}$$

the first-order optimality condition of Eq. (9) becomes $\mathbf{A}^T \mathbf{R}_{\Phi} = \mathbf{0}$.

The present implementation uses a Gauss–Newton approximation to obtain the reduced-state correction. Given the current reduced coordinates $\mathbf{a}^{(l)}$, the correction is determined from the linearized residual,

$$\Delta \mathbf{a}^{(l)} = \arg\min_{\Delta \mathbf{a} \in \mathbb{R}^r} \left\| \mathbf{R}_{\Phi}(\mathbf{a}^{(l)}, \boldsymbol{\mu}) + \mathbf{A}^{(l)} \Delta \mathbf{a} \right\|_2^2 \tag{11}$$

Here, $\mathbf{A}^{(l)}$ denotes the reduced-coordinate Jacobian defined in Eq. (10), evaluated at the current iterate $\mathbf{a}^{(l)}$, where $l$ is the Gauss–Newton iteration index. The reduced coordinates are updated as $\mathbf{a}^{(l+1)} = \mathbf{a}^{(l)} + \Delta \mathbf{a}^{(l)}$. In the full-residual implementation, only one Gauss–Newton correction is performed at each local design point, yielding an approximate solution to the least-squares problem in Eq. (9). The reduced-coordinate Jacobian can be constructed through residual directional evaluations without assembling the full state Jacobian, thereby reducing storage requirements and simplifying its construction.

The Galerkin formulation in Eq. (8) is retained for comparison with LSPG. Both formulations use the same reduced trial space, allowing the numerical experiments to examine the effect of residual projection independently of the state-space reduction.

### 2.2.2 Gradient evaluation with the reduced-order model

For the state representation in Eq. (7), define the reduced objective as $\hat{J}(\mathbf{a}, \boldsymbol{\mu}) = J(\bar{\mathbf{y}} + \boldsymbol{\Phi}\mathbf{a}, \boldsymbol{\mu})$. The reference state and basis are held fixed when evaluating derivatives within the current reduced space.

For the Galerkin formulation, the reduced adjoint variable $\boldsymbol{\eta} \in \mathbb{R}^r$ satisfies

$$\left( \frac{\partial \mathbf{R}_G}{\partial \mathbf{a}} \right)^T \boldsymbol{\eta} = -\left( \frac{\partial \hat{J}}{\partial \mathbf{a}} \right)^T \tag{12}$$

The corresponding design gradient is

$$\frac{d\hat{J}}{d\boldsymbol{\mu}} = \frac{\partial \hat{J}}{\partial \boldsymbol{\mu}} + \boldsymbol{\eta}^T \frac{\partial \mathbf{R}_G}{\partial \boldsymbol{\mu}} \tag{13}$$

where $\partial \mathbf{R}_G / \partial \mathbf{a} = \boldsymbol{\Phi}^T (\partial \mathbf{R} / \partial \mathbf{y}) \boldsymbol{\Phi}$ and $\partial \mathbf{R}_G / \partial \boldsymbol{\mu} = \boldsymbol{\Phi}^T (\partial \mathbf{R} / \partial \boldsymbol{\mu})$.

For LSPG, the gradient evaluation is based on a Gauss–Newton approximation to the linearization of the LSPG optimality condition. Specifically, terms involving derivatives of the reduced-coordinate Jacobian weighted by the residual are neglected. Under this approximation, a reduced dual variable $\boldsymbol{\psi} \in \mathbb{R}^r$ is introduced through

$$\mathbf{A}^T\mathbf{A}\boldsymbol{\psi} = -\left(\frac{\partial \hat{J}}{\partial \mathbf{a}}\right)^T \tag{14}$$

Defining $\boldsymbol{\lambda}_R = \mathbf{A}\boldsymbol{\psi}$, the corresponding gradient approximation becomes

$$\frac{d\hat{J}}{d\boldsymbol{\mu}} \approx \frac{\partial \hat{J}}{\partial \boldsymbol{\mu}} + \boldsymbol{\lambda}_R^T \frac{\partial \mathbf{R}}{\partial \boldsymbol{\mu}} \tag{15}$$

The residual-space dual variable therefore satisfies $\mathbf{A}^T\boldsymbol{\lambda}_R = -(\partial \hat{J} / \partial \mathbf{a})^T$. Since $r \ll N_R$, this system is underdetermined. When $\mathbf{A}$ has full column rank, $\boldsymbol{\lambda}_R = \mathbf{A}\boldsymbol{\psi}$ gives its minimum-norm solution. In the present implementation, the minimum-norm least-squares solution is computed directly using a stable least-squares solver, without explicitly forming the Moore–Penrose pseudoinverse. The resulting gradient is a Gauss–Newton-based approximation rather than the exact derivative of the single-step reduced-state update, and avoids solving a full-dimensional adjoint system.

### 2.3 Reduced-order Trajectory-based Optimization (RETO)

The preceding formulations provide reduced-order state and gradient evaluations for a prescribed trial space. RETO incorporates these calculations into an online optimization framework in which the reduced space is reconstructed from intermediate states generated during high-fidelity primal iterations.

Let $\boldsymbol{\mu}_k$ denote the design variables at the beginning of online stage $k$, and let $\mathbf{y}_k^{(0)}$ denote the initial state for the corresponding primal advancement. The high-fidelity solver is advanced for $n_k$ iterations, producing states $\mathbf{y}_k^{(0)}, \mathbf{y}_k^{(1)}, \ldots, \mathbf{y}_k^{(n_k)}$. Snapshots sampled from this trajectory are combined with previously retained states to form

$$\mathbf{S}_k = \left[\mathbf{y}_k^{(\tau_{k,1})}, \ldots, \mathbf{y}_k^{(\tau_{k,m_k})}, \mathbf{H}_k\right] \in \mathbb{R}^{N_y \times N_{s,k}} \tag{16}$$

Here, $\tau_{k,1}, \ldots, \tau_{k,m_k}$ are the selected primal iteration indices, $m_k$ is the number of snapshots collected from the current trajectory, $\mathbf{H}_k$ contains historical snapshots retained from recent online stages, and $N_{s,k}$ denotes the total number of snapshots in $\mathbf{S}_k$. The snapshots are not required to satisfy the high-fidelity equations to convergence. The current trajectory provides information about the state variations at the latest design, while the historical snapshots preserve information from previously visited designs.

The local reduced space is constructed by centering the snapshot matrix and applying proper orthogonal decomposition,

$$
\begin{aligned}
\bar{\mathbf{y}}_k &= \frac{1}{N_{s,k}} \sum_{i=1}^{N_{s,k}} \mathbf{s}_{k,i} \\
\mathbf{X}_k &= \left[ \mathbf{s}_{k,1} - \bar{\mathbf{y}}_k, \ldots, \mathbf{s}_{k,N_{s,k}} - \bar{\mathbf{y}}_k \right] \\
\mathbf{X}_k &= \mathbf{U}_k \boldsymbol{\Sigma}_k \mathbf{V}_k^T \\
\boldsymbol{\Phi}_k &= \mathbf{U}_k(:, 1:r_k)
\end{aligned} \tag{17}
$$

Here, $\boldsymbol{s}_{k,i}$ denotes the $i$th column of $\boldsymbol{S}_k$. The reduced basis $\boldsymbol{\Phi}_k$ is constructed by retaining the first $r_k$ columns of $\mathbf{U}_k$, where $r_k$ denotes the number of retained POD modes. The resulting reference state and reduced basis remain fixed throughout the subsequent local optimization stage.

Starting from $\boldsymbol{\mu}_{k,0} = \boldsymbol{\mu}_k$, let $\boldsymbol{\mu}_{k,j}$ denote the design variables after $j$ local optimization steps. The reduced-state correction and adjoint gradient described in Section 2.2 are used to update the design variables within the current reduced space. The objective evaluated within the current reduced space can be expressed as

$$
J_{loc}(\boldsymbol{\mu}) = J\left(\bar{\mathbf{y}}_k + \boldsymbol{\Phi}_k \mathbf{a}_k(\boldsymbol{\mu}), \boldsymbol{\mu}\right) + \rho_k(\boldsymbol{\mu}) \tag{18}
$$

Here, $a_k(\boldsymbol{\mu})$ denotes the reduced coordinates obtained at the current design using the state solution procedure in Section 2.2, and $\rho_k(\boldsymbol{\mu})$ represents the local regularization term, which is set to zero when no additional regularization is applied. The corresponding design gradient is evaluated using the reduced-order adjoint approximation, with the derivative of the regularization term included when applicable. The same reduced space supports up to $N_{\text{opt,max}}$ local design updates without further high-fidelity primal advancement.

Let $N_{\text{opt},k} \le N_{\text{opt,max}}$ denote the number of local optimization steps actually performed during stage $k$. The final design and its reconstructed reduced state are then transferred to the next online stage:

$$
\begin{aligned}
\boldsymbol{\mu}_{k+1} &= \boldsymbol{\mu}_{k,N_{\text{opt},k}} \\
\mathbf{y}_{k+1}^{(0)} &= \bar{\mathbf{y}}_k + \boldsymbol{\Phi}_k \mathbf{a}_k\left(\boldsymbol{\mu}_{k,N_{\text{opt},k}}\right)
\end{aligned} \tag{19}
$$

A new high-fidelity trajectory is then generated at the updated design, and the local reduced space is reconstructed using the current and retained historical snapshots. The procedure alternates between trajectory acquisition and reduced-order optimization throughout the design process.

By constructing reduced spaces from intermediate primal states and reusing each local model for multiple design updates, RETO reduces the frequency of high-fidelity computations and avoids requiring a converged full-order solution at every model update. In the present implementation, the number of primal iterations per stage, snapshot sampling interval, amount of retained historical data, and maximum number of local optimization steps are prescribed. The complete procedure is illustrated in Fig. 1.

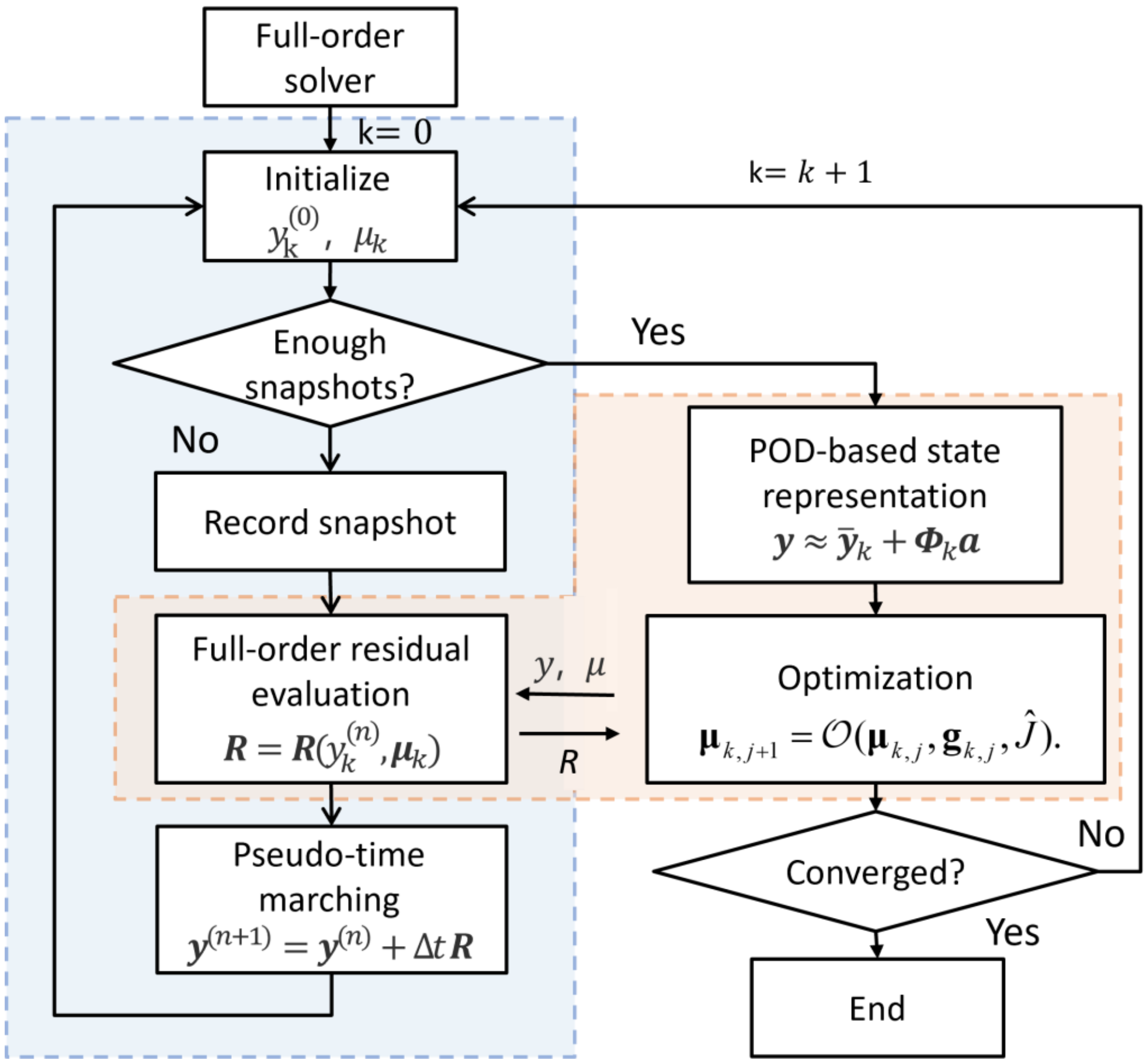

Fig. 1. Overall workflow of the RETO online reduced-order adjoint optimization framework.

## 3. Numerical experiments

This section evaluates RETO on three PDE-constrained optimization problems of increasing complexity: Poisson parameter field inversion, logistic resource allocation, and RANS airfoil pressure inverse design. The Poisson and logistic problems examine the method for parameter inversion and constrained distributed optimization, respectively, while the RANS problem further assesses its performance for large-scale nonlinear flows under both subsonic and transonic conditions. The traditional direct adjoint loop (DAL) is used as the primary reference. To avoid assessing RETO solely from its reduced-order predictions, selected designs along the RETO optimization trajectory are independently evaluated using high-fidelity simulations. These simulations are used only to determine the corresponding high-fidelity states and objective values and do not participate in reduced-space construction, gradient evaluation, or design updates.

Computational efficiency is measured by the main optimization wall-clock time required to reach a common target level. For each case, the common threshold is determined a posteriori from the optimization histories and is used only for time-to-threshold comparison; it does not affect online model updates or optimization termination. The time-to-threshold speedup is defined as

$$S_{th} = \frac{t_{DAL}}{t_{RETO}} \tag{20}$$

where $t_{DAL}$ and $t_{RETO}$ denote the main optimization times required by DAL and RETO, respectively, to reach the same evaluation threshold. The Poisson and Logistic cases are both implemented in Python on a 64-bit Windows platform and run in a single Python process. Their main RETO settings are summarized in Table 2, with DAL used as the reference in both cases.

Table 2. RETO algorithmic settings for the Poisson and Logistic cases

| **Algorithmic setting** | **Cases** | |
|---|---|---|
| | **Poisson** | **Logistic** |
| Optimizer | CG | L-BFGS-B |
| High-fidelity steps per stage, $N_{\mathrm{HF}}$ | 6750 | 4500 |
| Snapshot sampling interval, $N_{\mathrm{samp}}$ | 150 | 100 |
| Current snapshots, $N_{\mathrm{cur}}$ | 45 | 45 |
| Historical snapshots retained, $N_{\mathrm{hist}}$ | 15 | 15 |
| POD dimension, $r$ | 16 (fixed) | 16 (fixed) |
| Maximum local optimization steps, $N_{\mathrm{opt,max}}$ | 10 | 30 |

## 3.1 Poisson parameter field inversion

We first consider the inverse problem of recovering a spatially varying diffusion coefficient in a two-dimensional Poisson equation. This problem is used as a baseline test of RETO for distributed PDE-constrained optimization, in which the design variable is a spatially varying field rather than a small set of scalar parameters. The objective is to reconstruct the diffusion coefficient from observations of the corresponding state field. This setting provides a simple but representative test of whether the local reduced model constructed from unconverged primal trajectories can support repeated state and gradient evaluations during a distributed inverse problem. The governing equation is

$$\nabla \cdot (D \nabla u) + f = 0, \quad (x, y) \in \Omega \tag{21}$$

The domain is $\Omega = [-1,1] \times [-1,1]$, where the boundary condition is $u = 0.01\cos(\pi x)\cos(\pi y)$ and the source term is $f = \sin(\pi x)\cos(\pi y)$. The target state $u_{target}$ is generated from the full-order solution corresponding to the reference diffusion coefficient $D_{true} = \sqrt{2} + 0.1(y - x)$, and the inversion is initialized with $D_0(x, y) = 0.8 + 0.25(y - x) + 0.15\sin(2\pi x)\sin(2\pi y)$. The objective function consists of the mean squared state error at the observation points and a weak smoothing regularization term,

$$J(D) = \frac{1}{N_{obs}} \sum_{i=1}^{N_{obs}} (u_i - u_{target,i})^2 + \alpha_D (\| \nabla_x D \|_2^2 + \| \nabla_y D \|_2^2) \tag{22}$$

Here, $N_{obs} = 100$ and $\alpha_D = 1.0 \times 10^{-7}$. A $51 \times 51$ uniform grid is used. Observation points are placed regularly in the interior of the domain at intervals of five grid nodes in both the $x$ and $y$ directions, giving $10 \times 10 = 100$ observation points. RETO and DAL use the same discretization, initial parameter field, observation locations, and objective function.

Figure 2 compares the reference fields with the final inversion results obtained by RETO and DAL. Both methods recover the main spatial features of the reference diffusion coefficient and produce state fields close to the target. The relative errors in the diffusion coefficient are 0.7591% for RETO and 0.7470% for DAL, while the corresponding state errors are 0.1352% and 0.1470%, respectively. These results indicate that RETO achieves final inversion accuracy comparable to that of the full-order adjoint optimization.

Figure 3 further compares the spatial errors and optimization histories. The plotted quantity is the observation mismatch $J_{obs}$, excluding the regularization term in Eq. (22). Across the 50 independent high-fidelity validation points, the validated objective values closely track the reduced-order prediction throughout the convergence process. Using $J_{obs} = 10^{-9}$ as the common evaluation threshold, the high-fidelity validation of RETO reaches this level at 52.49 s, compared with 491.06 s for DAL, corresponding to a speedup of 9.36×.

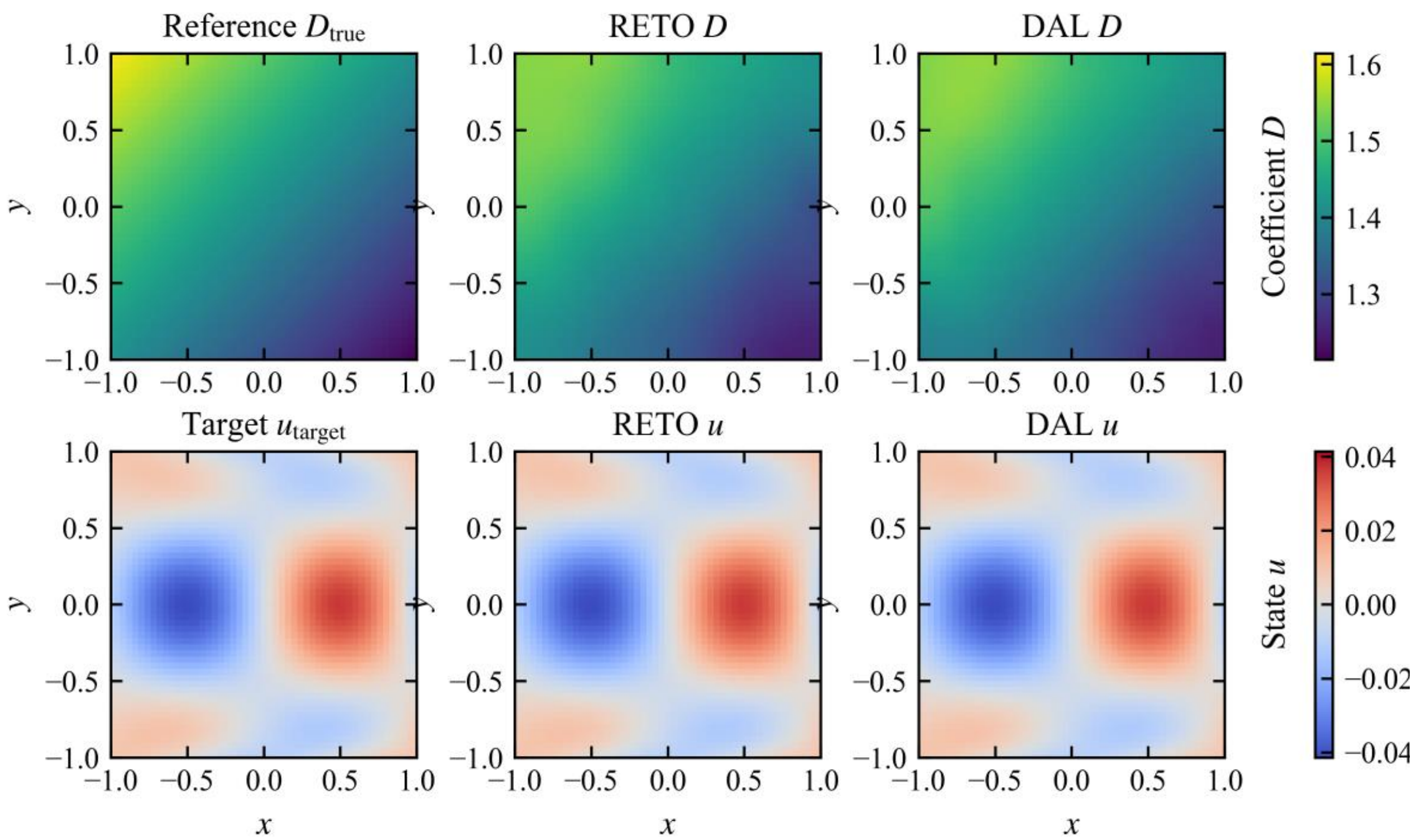

Fig. 2. Reference and reconstructed fields for the Poisson parameter inversion.

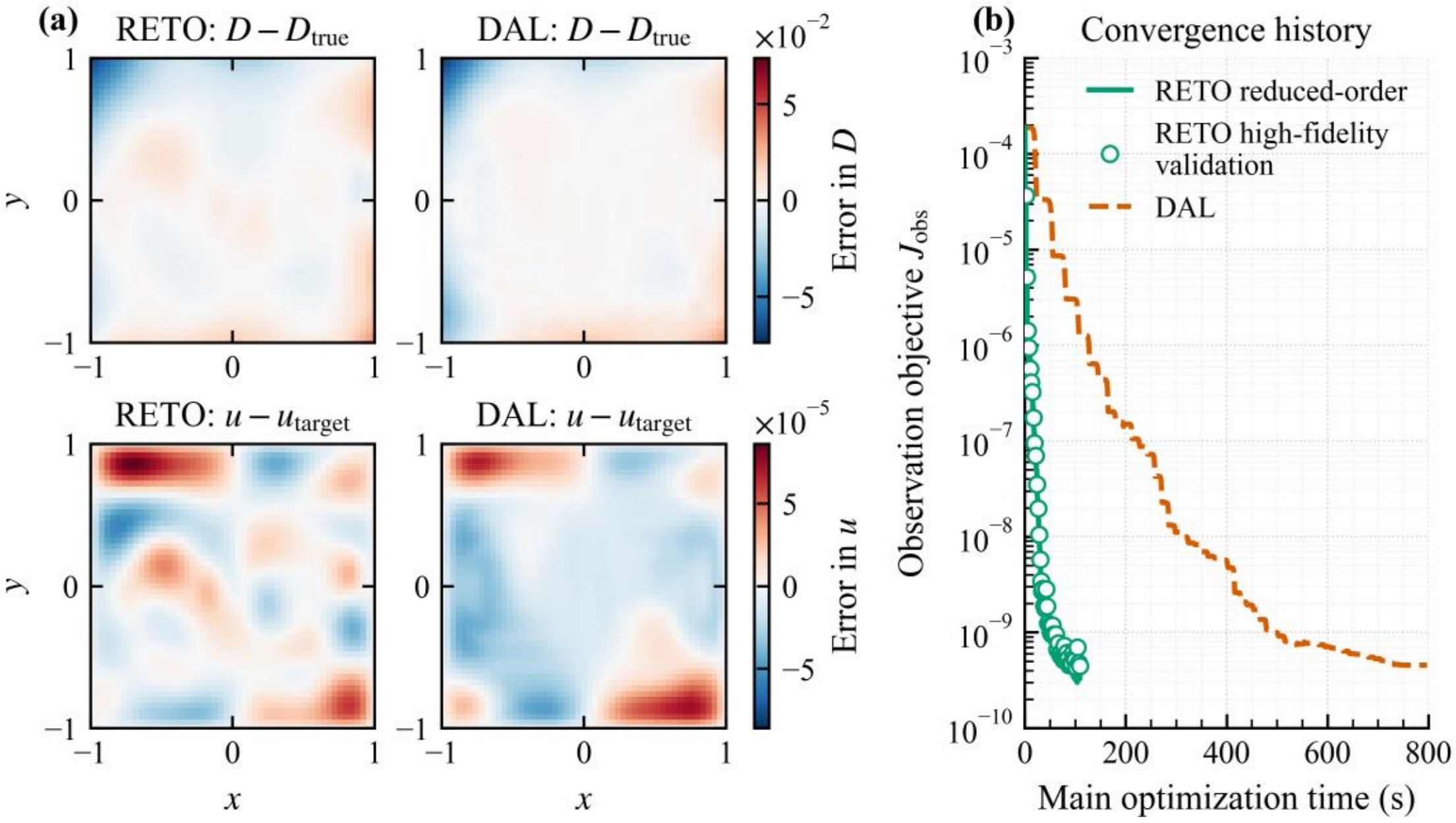

Fig. 3. Poisson parameter inversion results: (a) spatial errors in the reconstructed diffusion coefficient and state for RETO and DAL; (b) observation-mismatch histories for RETO reduced-order prediction, RETO high-fidelity validation, and DAL.

### 3.2 Logistic resource allocation

The second model problem considers two-dimensional resource allocation governed by a diffusive logistic equation. In contrast to the preceding parameter-inversion problem, this case represents a nonlinear constrained distributed optimization problem. The design variable is the spatial resource distribution $m(x,y)$, while the state $u(x,y)$ describes the corresponding steady population response. The optimization seeks a resource distribution that balances population growth against resource cost under both local bounds and a prescribed total-resource constraint. This problem therefore provides a further test of RETO in the presence of nonlinear state dynamics, distributed design variables, and explicit feasibility constraints.

The state variable $u(x,y)$ satisfies

$$\kappa\Delta u + mu - u^2 = 0, \quad (x,y)\in\Omega \tag{23}$$

where $\Omega=[0,1]\times[0,1]$, with homogeneous Neumann boundary conditions. The resource distribution is subject to pointwise bounds together with a prescribed total-resource constraint,

$$0 \le m(x,y) \le 1, \quad \frac{1}{|\Omega|}\int_\Omega m(x,y)d\Omega = \delta \tag{24}$$

Accounting for both the steady total population benefit and the cost of resource allocation, the benefit function is defined as

$$B(u,m)=\frac{1}{|\Omega|}\int_{\Omega}\left(u-\beta m^{2}\right)d\Omega \tag{25}$$

Previous studies have shown that optimal resource distributions in related logistic models can exhibit pronounced concentration and bang-bang structures under certain conditions [22]. The steady-state logistic model and quadratic resource-cost functional adopted in the present study are based on the formulation of Ding et al. [23]. Here, $\kappa=0.1$, $\delta=0.5$, $\beta=0.1$, and a $41\times41$ uniform grid is used. The initial resource field is $m_0(x,y)=0.5+0.15(0.5-x)$. This slight asymmetry is introduced to avoid convergence toward a symmetric solution associated with the symmetry of the square domain. The numerical implementation uses L-BFGS-B [24] with $-B$ as the minimization objective. Because L-BFGS-B directly handles only pointwise bounds, the total-resource constraint is enforced through projection onto the feasible set. RETO and DAL use the same projection and gradient treatment and therefore operate within the same feasible region.

Figure 4 compares the final state and resource distributions obtained by RETO and DAL. Both methods produce a strongly concentrated resource distribution with a narrow switching region between resource-rich and resource-poor areas, consistent with the characteristic bang-bang structure of this problem. The relative differences between RETO and DAL are 0.1828% for the state field and 3.8327% for the resource field. The larger difference in $m$ is concentrated mainly near the switching interface, whereas most of the saturated regions remain nearly unchanged. The much smaller difference in the state field indicates that these localized variations in the resource distribution have only a limited effect on the resulting population response. Both methods also satisfy the prescribed resource constraints to numerical precision.

Figure 5 compares the RETO reduced-order benefit, the independent high-fidelity validation, and the DAL result as functions of main optimization time. The reduced-order prediction exhibits a transient overshoot during the early stage, but this discrepancy is rapidly reduced as new high-fidelity trajectory information is incorporated and the local ROM is rebuilt. The high-fidelity validation subsequently remains close to the reduced prediction and approaches the benefit level obtained by DAL. Using $B=0.5337$ as the common evaluation threshold, the high-fidelity validation of RETO reaches this level at 0.812 s, compared with 18.170 s for DAL, corresponding to a time-to-threshold speedup of 22.38×. These results show that RETO can retain the dominant optimal resource-allocation structure and the associated nonlinear state response while substantially reducing the computational cost of the constrained optimization.

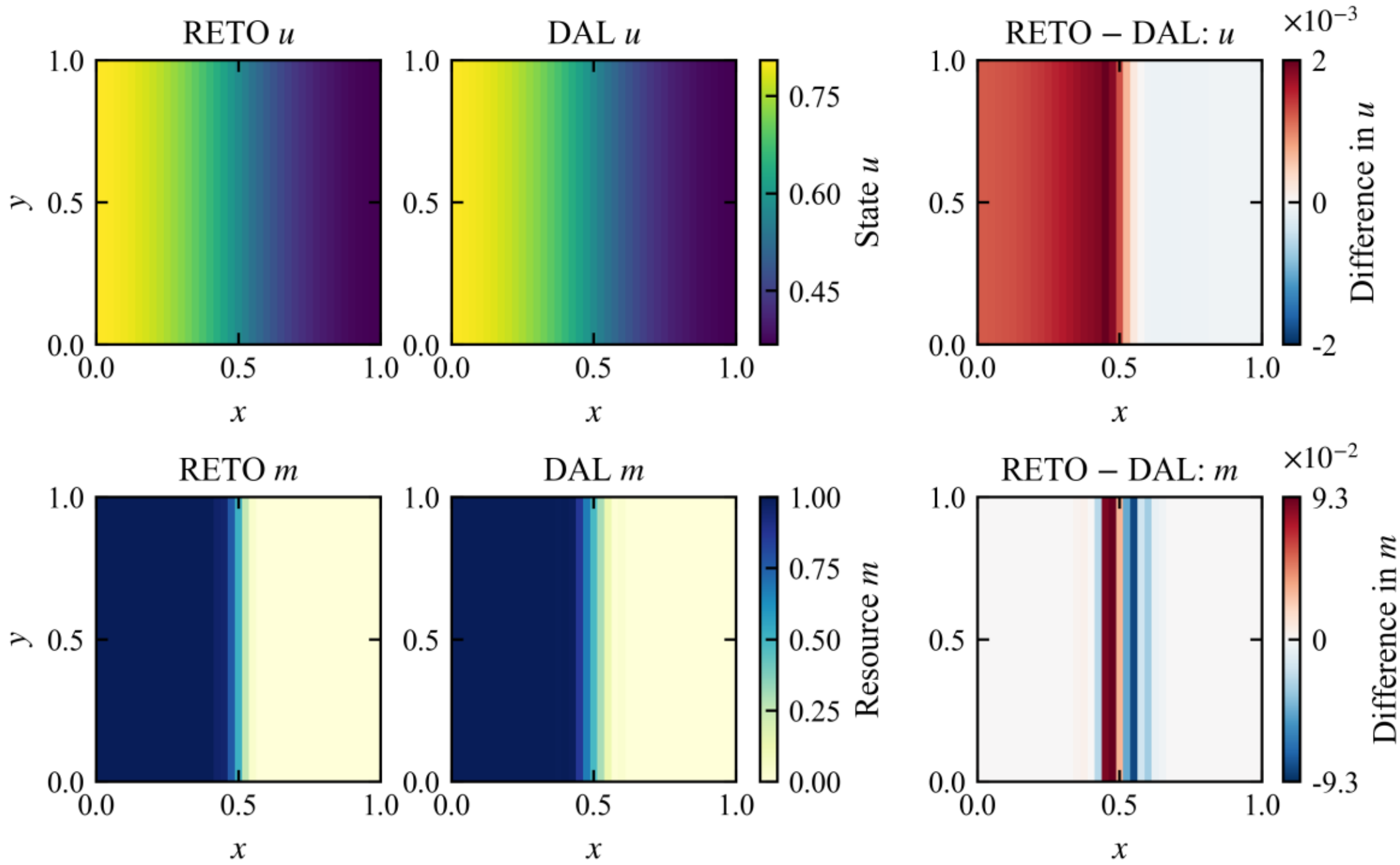

Fig. 4. Final states and resource distributions for the logistic resource allocation problem.

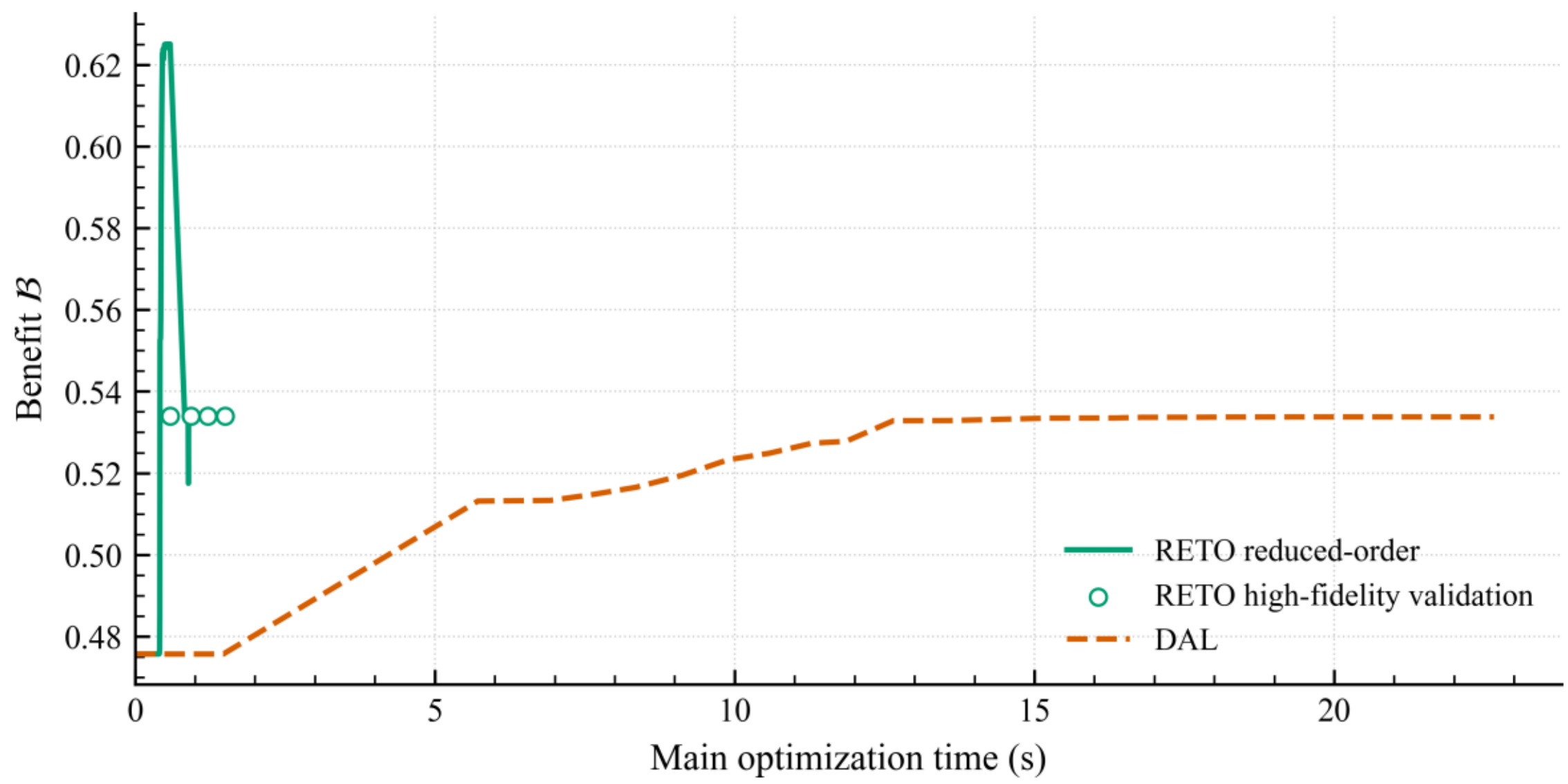

Fig. 5. Optimization histories for the logistic resource allocation problem.

### 3.3 RANS airfoil pressure inverse design

We next consider RANS airfoil pressure inverse design to evaluate RETO for high-dimensional nonlinear flow optimization. Compared with the preceding problems, RANS flows involve a much larger state dimension and stronger nonlinear coupling. Under transonic conditions, geometric changes also alter the shock position and strength, increasing the difficulty of constructing an effective local low-dimensional representation [25]. The subsonic shock-free case is first used to compare the full-residual and projected-residual formulations, followed by the transonic shock-containing case to assess the inverse-design performance of RETO and the adaptation of the online local model along the design trajectory.

### 3.3.1 Problem and numerical setup

The airfoil pressure inverse-design problem seeks an airfoil geometry that reproduces a prescribed wall-pressure distribution. NACA0012 is used as the initial airfoil, while the full-order RANS pressure distribution of RAE2822 under the corresponding freestream conditions is taken as the target. Two flow conditions are considered. The subsonic case is defined by $M_\infty = 0.3$, an angle of attack of $2^\circ$, and $\mathrm{Re} = 6.0\times10^6$, whereas the transonic case uses $M_\infty = 0.734$, an angle of attack of $2.79^\circ$, and $\mathrm{Re} = 6.5\times10^6$. The Spalart–Allmaras (SA) turbulence model [26] is employed, and the flow is computed on a quasi-two-dimensional mesh formed by a single-layer extrusion.

The steady RANS equations are discretized using a cell-centered finite-volume method with second-order Green–Gauss reconstruction for variable gradients. The inviscid flux is evaluated using Roe's approximate Riemann solver [27], with an entropy fix applied to the acoustic eigenvalues, and the viscous flux is computed using a weighted-neighbor formulation. No-slip adiabatic conditions are imposed on the airfoil surface, while a non-reflecting characteristic boundary condition based on Riemann invariants is applied at the far field. The equations are advanced using Jacobian-free symmetric Gauss–Seidel (Sym-GS) implicit pseudo-time marching with a CFL number of 8. The airfoil geometry is parameterized using a fifth-order class-shape transformation (CST) representation [28], with six Bernstein coefficients on each surface, giving 12 design variables. The computational mesh contains 39,402 cells.

For RETO, the snapshot state consists of density, the two in-plane momentum components, and total energy at each cell center, resulting in a state dimension of 157,608. The SA working variable is excluded from the POD state vector. It is updated together with the RANS equations during high-fidelity pseudo-time advancement and snapshot acquisition, but is frozen at the value obtained at the end of the latest high-fidelity advancement during subsequent reduced-state correction, residual directional differentiation, and gradient evaluation. Accordingly, the term *full residual* refers here to the unprojected RANS discrete residual of the four conservative variables under the frozen-SA treatment.

The inverse-design objective is defined as the mean squared pressure-coefficient error over $N_w = 200$ wall sampling points,

$$J_p(\boldsymbol{\mu}) = \frac{1}{N_w}\sum_{q=1}^{N_w}[C_{p,q}(\boldsymbol{\mu}) - C_{p,q}^{target}]^2 \tag{26}$$

To limit the design variation within each online model stage, the local optimization uses

$$J_{loc} = J_p + \frac{\gamma}{n_\mu}\|\boldsymbol{\mu} - \boldsymbol{\mu}_k\|_2^2, n_{\boldsymbol{\mu}} = 12 \quad (27)$$

The regularization term is used only to control design variation during the current local optimization and is not included in the final assessment of inverse-design quality. All final values, convergence histories, and time-to-threshold comparisons reported below are therefore based on $J_p$. The design variables are updated using SLSQP [29]. The main numerical and RETO settings for the two flow conditions are summarized in Table 3.

Table 3. Main numerical and RETO settings for the RANS inverse-design cases

| **Setting** | **Cases** | |
|---|---|---|
| | **Subsonic** | **Transonic** |
| $M_\infty$ | 0.3 | 0.734 |
| $Re$ | $6.0 \times 10^6$ | $6.5 \times 10^6$ |
| Angle of attack | 2° | 2.79° |
| Snapshot interval | 40 | 40 |
| Current snapshots | 20 | 75 |
| Historical batches | 3 | 5 |
| $r_{\max}$ | 30 | 75 |
| $\gamma$ | 0.05 | 1.0 |
| $N_{opt,\max}$ | 35 | 35 |

For both cases, the POD dimension is determined by $r = \min(\lfloor N_s / 3 \rfloor, r_{\max})$. Residual directional derivatives with respect to the reduced coordinates and design variables are evaluated using one-sided finite differences with $\varepsilon(z) = \max(|z|10^{-3}, 10^{-8})$. Except for the projected-residual comparison in the subsonic case, all RANS inverse-design calculations use the full-residual formulation. The reduced-coordinate residual Jacobian is re-evaluated whenever the design variables and geometry are updated. DAL uses the same mesh, flow discretization, CST parameterization, optimizer, and frozen-SA linearization treatment as RETO. All RANS calculations are performed on an Intel Core Ultra 9 285K platform using at most 20 CPU threads. The reported RETO main optimization time includes high-fidelity primal advancement, snapshot processing and POD reconstruction, Jacobian assembly, state and gradient evaluation, and optimizer updates. Independent high-fidelity validation is excluded from the reported main optimization time.

### 3.3.2 Subsonic case: residual treatment and inverse-design results

The subsonic shock-free case is first used to compare the full-residual and projected-residual formulations. Both use the same POD state variables, snapshot strategy, reduced dimension, and optimizer. The projected-residual formulation further compresses the full-order residual into a low-dimensional space, whereas the full-residual formulation reduces only the state-search dimension and determines the reduced state through least-squares correction. To examine the sensitivity of the projected formulation to local regularization, $\gamma = 0.05$, 10, 20, and 30 are considered, while the full-residual formulation uses $\gamma = 0.05$.

Figure 6 reports the mean absolute surface-coordinate error between the current and target airfoils as a measure of geometric convergence. This quantity is used only for a posteriori assessment and does not enter the optimization. With the projected-residual formulation and $\gamma = 0.05$, the shape error increases rapidly during the early stage and stable convergence is not achieved. Increasing $\gamma$ improves stability but slows the design evolution. By contrast, the full-residual formulation remains stable with $\gamma = 0.05$ and continuously reduces the shape error to approximately $10^{-5}$.

Figure 7 compares the final airfoils and wall-pressure distributions obtained by full-residual RETO, DAL, and the stable projected-residual case with $\gamma = 30$. The final geometry obtained by full-residual RETO nearly overlaps that of DAL, and the corresponding $C_p$ distributions remain close over most of the chord. The projected-residual formulation remains stable under stronger regularization, but its final geometry and pressure response still differ noticeably from DAL. The stronger regularization required for stability therefore also restricts the design evolution.

Figure 8 shows the pressure-mismatch objective $J_p$ of full-residual RETO and DAL as a function of main optimization time. The RETO reduced prediction and independent high-fidelity validation exhibit broadly consistent decreasing trends. Using $J_{\mathrm{p}} = 5\times10^{-7}$ as the common near-final threshold, RETO first reaches this level at **802.7 s**, whereas DAL requires **6904.8 s**, corresponding to a time-to-threshold speedup of **8.60×**. The full-residual formulation is therefore used in the subsequent transonic case.

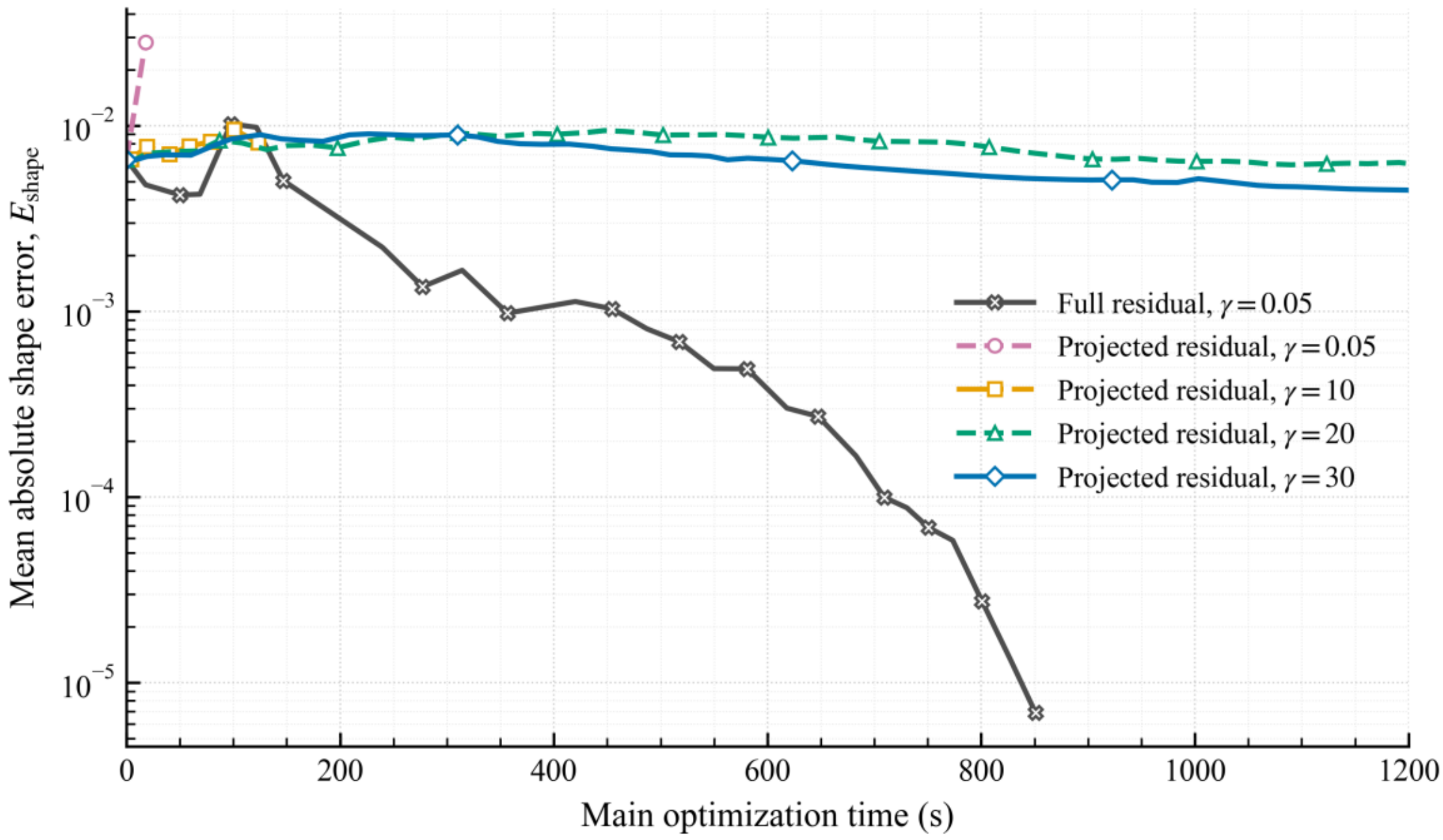

Fig. 6. Effect of residual treatment and regularization on shape-error convergence in the subsonic case.

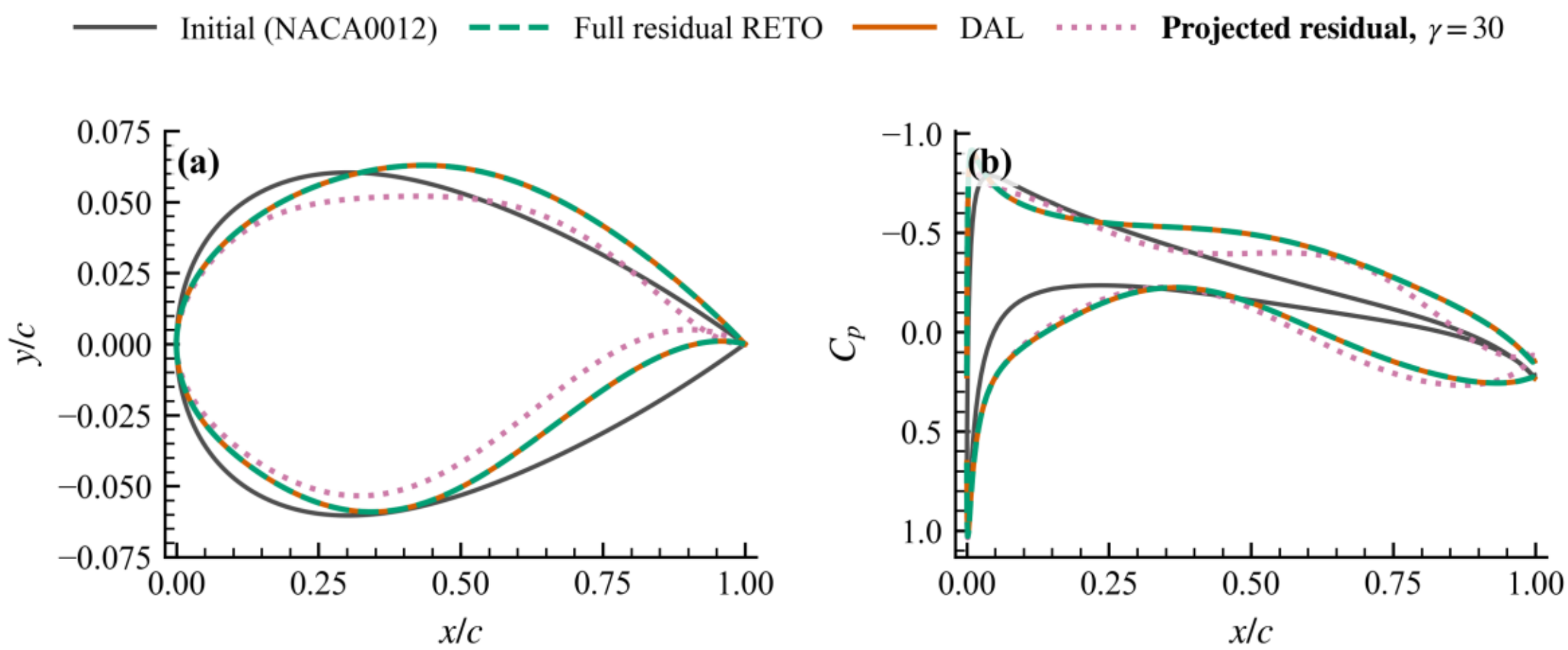

Fig. 7. Subsonic inverse-design results: (a) final airfoil geometries; (b) wall-pressure distributions.

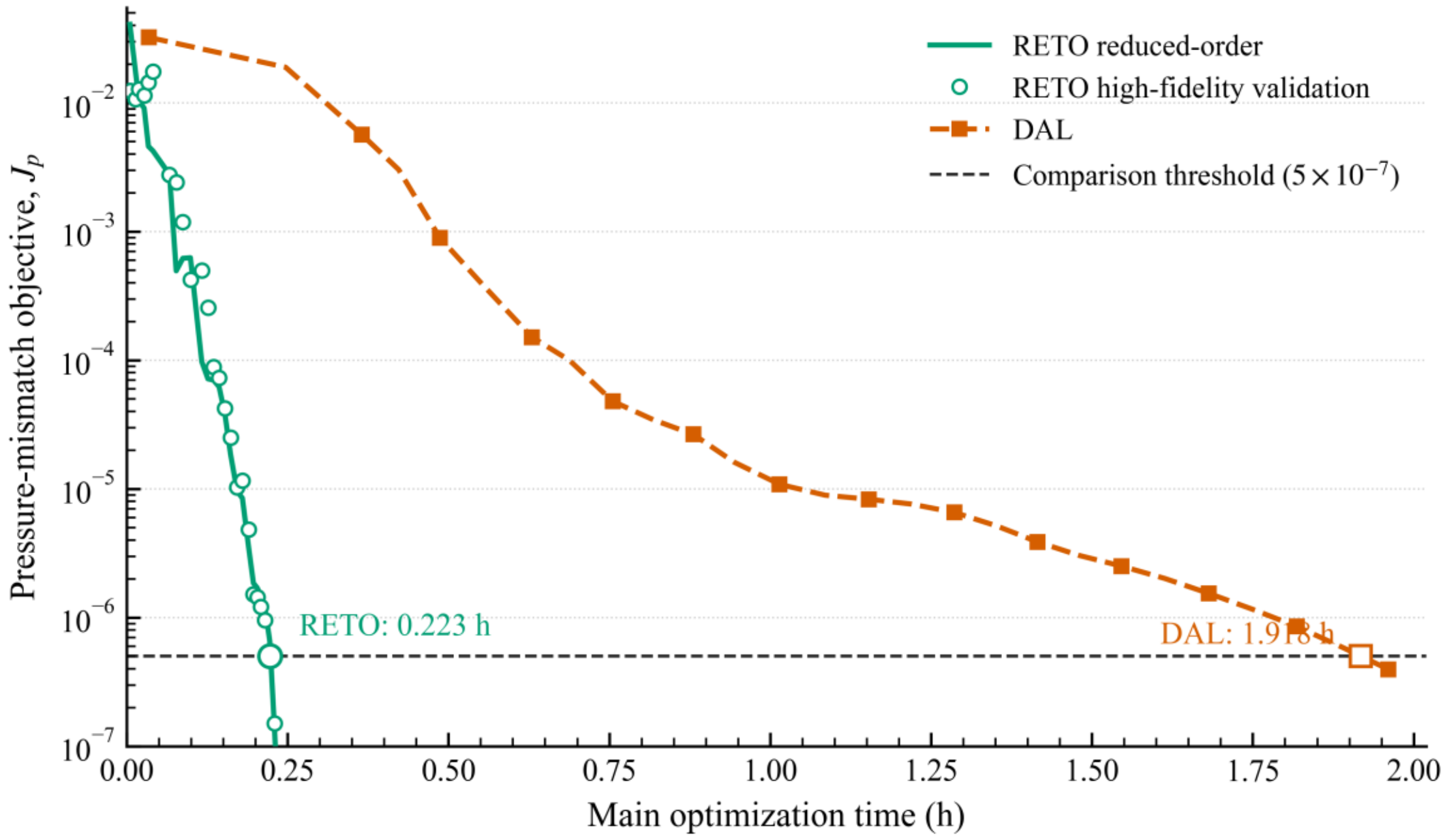

Fig. 8. Pressure-mismatch objective histories for the subsonic inverse-design case.

### 3.3.3 Transonic case: inverse design and online model adaptation

Compared with the subsonic flow, geometric changes alter both the location and strength of the shock, making it more difficult for the local state space to represent the current design neighborhood accurately [25]. Figure 9 compares the final airfoils and wall-pressure distributions obtained by RETO and DAL. The two methods produce closely matching final geometries, and the corresponding pressure distributions reproduce the main transonic pressure features, including the response in the shock region.

Figure 10 compares the values of $J_p$ from the reduced-order prediction, independent high-fidelity validation, and DAL as functions of main optimization time. The transonic RETO calculation comprises 83 online model stages. During the early stages, a noticeable discrepancy is observed between the reduced prediction and the independent high-fidelity validation, and the reduced pressure error exhibits some non-monotonic behavior. As additional high-fidelity trajectories are incorporated, the reduced prediction becomes increasingly consistent with the high-fidelity validation as both decrease. Using $J_p = 10^{-6}$ as the common threshold, RETO first reaches this level at **5461.4 s** (approximately 1.52 h, stage 74), whereas DAL requires **15056.1 s** (approximately 4.18 h), corresponding to a time-to-threshold speedup of **2.76×**.

To quantify the agreement between the reduced prediction and high-fidelity validation, the error ratio is defined as $r_J = J_{\text{ROM}} / J_{\text{FOM}}$, and the relative difference as $\varepsilon_J = \frac{|J_{\text{ROM}} - J_{\text{FOM}}|}{|J_{\text{FOM}}|} \times 100\%$. For the 46 validation points within the first hour, the median values of $r_J$ and $\varepsilon_J$ are **0.784** and **26.3%**, respectively. For the subsequent 37 validation points, these values become **1.021** and **3.0%**, while the maximum $r_J$ is 1.092. At stage 2, the high-fidelity validation pressure error is approximately 4.89 times the corresponding reduced error, whereas by stage 60, $r_J$ has approached 1.031.

This evolution reflects the adaptation of the online local model along the design trajectory. At the beginning of the optimization, the snapshots are generated mainly from trajectories close to the initial airfoil and therefore contain limited information about later geometric changes and shock motion. As new unconverged iterative states are incorporated and combined with recent historical trajectories to reconstruct the POD space, the reconstructed local model progressively captures features of the current flow response and shock motion. The reduced prediction and high-fidelity validation therefore evolve from a clear early mismatch toward consistent behavior at later stages.

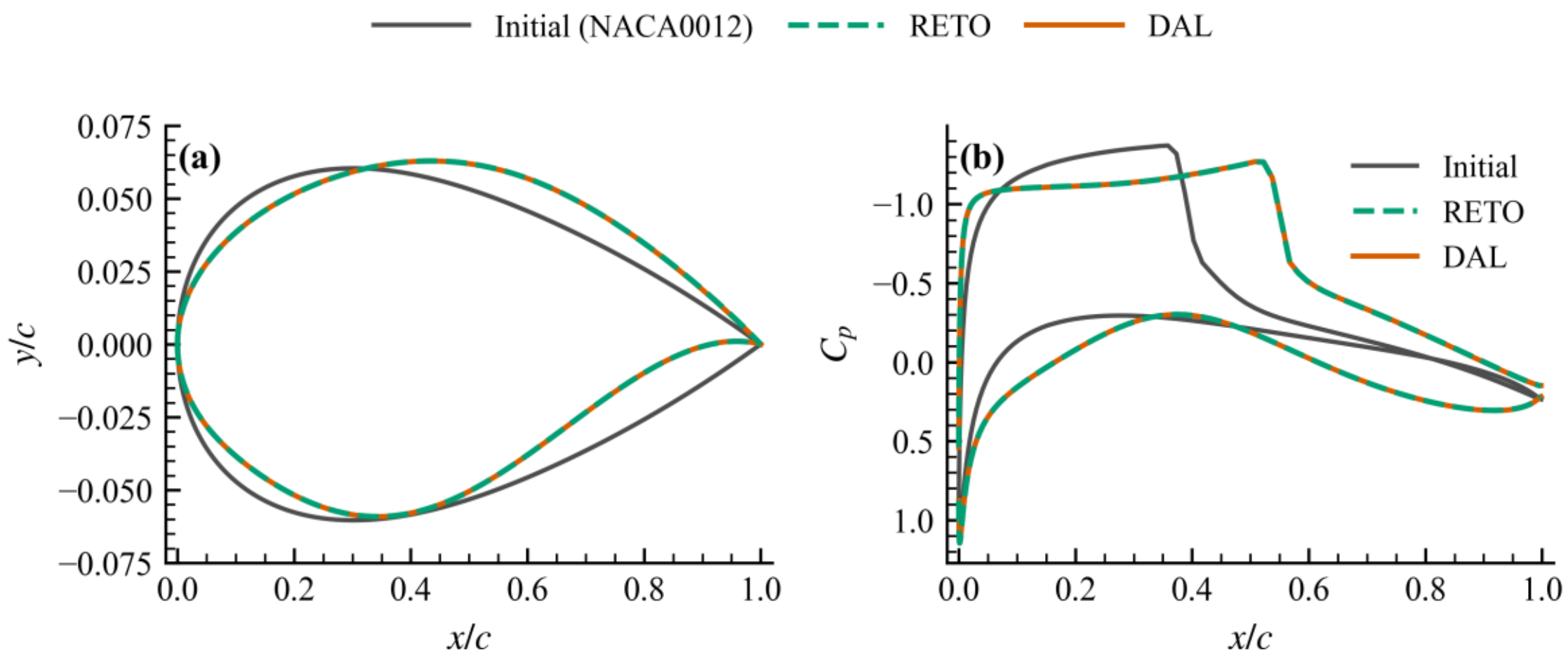

Fig. 9. Transonic inverse-design results: (a) final airfoil geometries; (b) wall-pressure distributions.

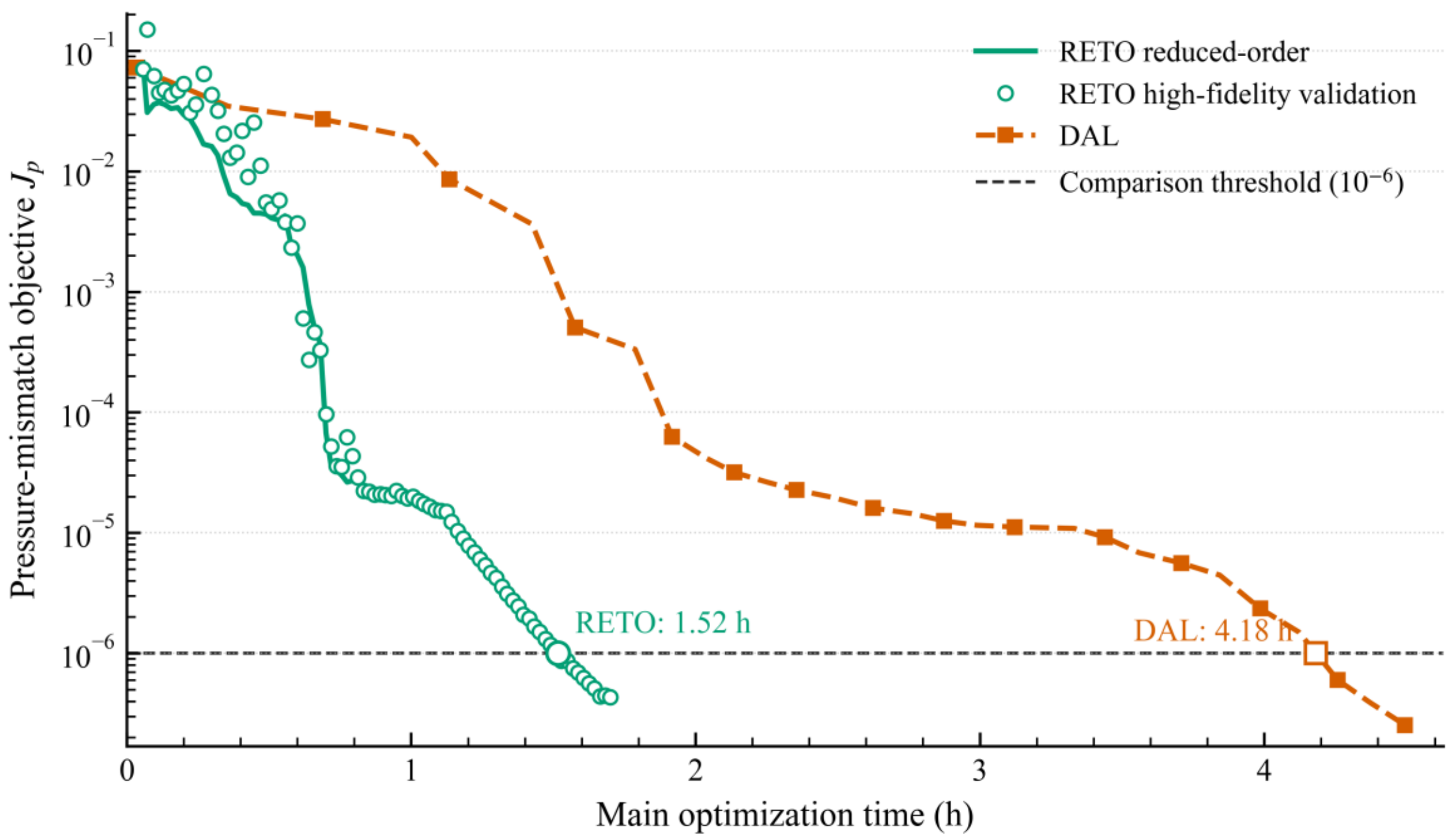

Fig. 10. Pressure-mismatch objective histories for the transonic inverse-design case.

### 3.3.4 Computational cost analysis

Table 4 summarizes the breakdown of the main RETO optimization time before the common threshold is reached. In the subsonic case, Jacobian assembly and reduced-order adjoint evaluation account for **36.16%** and **32.65%** of the total time, respectively, whereas high-fidelity snapshot acquisition accounts for **22.00%**. In the transonic case, the fraction associated with snapshot acquisition increases to **57.46%**, becoming the dominant computational cost, while Jacobian assembly and reduced-order adjoint evaluation account for **26.59%** and **6.18%**, respectively.

Table 4. Breakdown of RETO main optimization time before reaching the common threshold

| Timing component | Time fraction (%) | |
|---|---|---|
| | Subsonic | Transonic |
| High-fidelity snapshot acquisition | 22.00 | 57.46 |
| POD reconstruction | 1.94 | 7.29 |
| Jacobian assembly | 36.16 | 26.59 |
| Reduced-order adjoint evaluation | 32.65 | 6.18 |
| Other optimization overhead | 7.25 | 2.48 |
| **Total** | **100.00** | **100.00** |

## 4. Conclusions

This paper presents Reduced-order Trajectory-based Optimization (RETO), an online reduced-order optimization framework that constructs local POD bases from intermediate states generated during high-fidelity primal iterations. Unlike approaches that rely on converged high-fidelity solutions for model construction, RETO uses trajectories obtained from a finite number of primal iterations. Each local reduced model supports several consecutive state evaluations, adjoint-based gradient evaluations, and design updates before the basis is reconstructed using newly acquired trajectories and recent historical data. By reusing information from partially converged primal solutions, RETO reduces the need for repeated full-order convergence throughout the optimization process, without requiring a reduced basis that covers the entire design space in advance.

The method is evaluated through Poisson parameter field inversion, logistic resource allocation, and subsonic and transonic RANS airfoil pressure inverse design. In all four cases, RETO produces results comparable to those of the traditional direct adjoint loop (DAL), with time-to-threshold speedups of 9.36×, 22.38×, 8.60×, and 2.76×, respectively. The main RETO formulation uses a single Gauss–Newton correction based on the full discrete residual to update the reduced state, followed by reduced-order adjoint gradient evaluation. The subsonic RANS comparisons show that retaining the full residual leads to more stable optimization than further projecting it onto the reduced space. The influence of local regularization is also examined for the projected-residual formulation. Increasing the regularization coefficient improves stability but restricts design evolution, whereas the full-residual formulation achieves stable convergence with a smaller coefficient and produces airfoil geometries and pressure distributions close to those obtained with DAL.

The transonic results further demonstrate the benefit of reconstructing the local reduced space as optimization progresses. During the early stages, the reduced predictions show noticeable differences from independent high-fidelity evaluations. These differences decrease as new primal trajectories are incorporated into the model. The updated reduced space progressively represents the changes in the flow field, including variations in shock position and strength, and the final inverse-design results agree closely with those obtained using DAL. This behavior indicates that an effective reduced representation can be maintained along the optimization trajectory through repeated local reconstruction, without requiring accurate representation of the entire design space from the outset.

Several limitations warrant further investigation. The present aerodynamic examples focus on surface-pressure inverse design, and further studies are needed to assess the method for objectives involving other flow quantities, particularly near-wall quantities such as skin friction. The current implementation also relies on prescribed settings for trajectory acquisition, snapshot selection, reduced-space construction, and local optimization steps. Although independent high-fidelity validations demonstrate the effectiveness of RETO in the tested cases, the framework does not yet incorporate an online error estimator or a model-management mechanism to assess the reliability of local reduced-order predictions and guide optimization termination. Developing adaptive strategies based on information available during primal iterations and reduced-order optimization may therefore improve both computational efficiency and reliability. Further analysis of model accuracy and convergence is also needed to establish the conditions under which RETO can be applied reliably.